\documentclass{ws-fnl}
\usepackage{cite}

\begin{document}

\markboth{Q. Cai, H.-C. Xu \& W.-X. Zhou}{Taylor's Law of Temporal Fluctuation Scaling in Stock Illiquidity}

\catchline{}{}{}{}{}

\title{Taylor's Law of Temporal Fluctuation Scaling in Stock Illiquidity}

\author{\footnotesize QING CAI}
\address{Department of Finance, School of Busness\\
East China University of Science and Technology\\
Shanghai 200237, China}

\author{\footnotesize HAI-CHUAN XU}
\address{Department of Finance, School of Busness\\ Postdoctoral Research Station\\ Research Center for Econophysics\\
East China University of Science and Technology\\
Shanghai 200237, China}

\author{\footnotesize WEI-XING ZHOU\footnote{Corresponding author.}}
\address{Department of Finance, School of Busness\\ Department of Mathematics, School of Science\\ Research Center for Econophysics\\
East China University of Science and Technology\\
Shanghai 200237, China\\
wxzhou@ecust.edu.cn}

\maketitle

\begin{history}
\received{2 April 2016}
\revised{(revised date)}
\end{history}

\begin{abstract}
  Taylor's law of temporal fluctuation scaling, variance $\sim$ $a($mean$)^b$, is ubiquitous in natural and social sciences. We report for the first time convincing evidence of a solid temporal fluctuation scaling law in stock illiquidity by investigating the mean-variance relationship of the high-frequency illiquidity of almost all stocks traded on the Shanghai Stock Exchange (SHSE) and the Shenzhen Stock Exchange (SZSE) during the period from 1999 to 2011. Taylor's law holds for A-share markets (SZSE Main Board, SZSE Small \& Mediate Enterprise Board, SZSE Second Board, and SHSE Main Board) and B-share markets (SZSE B-share and SHSE B-share). We find that the scaling exponent $b$ is greater than 2 for the A-share markets and less than 2 for the B-share markets. We further unveil that Taylor's law holds for stocks in 17 industry categories, in 28 industrial sectors and in 31 provinces and direct-controlled municipalities with the majority of scaling exponents $b\in(2,3)$. We also investigate the $\Delta{t}$-min illiquidity and find that the scaling exponent $b(\Delta{t})$ increases logarithmically for small $\Delta{t}$ values and decreases fast to a stable level.

  \medskip
  \noindent{\textit{keywords}}: Taylor's law; mean-variance analysis; temporal fluctuation scaling; stock illiquidity; Chinese stock markets
\end{abstract}



\section{Introduction}

A complex system is composed of many interacting constituents labelled by indices $\{i|i=1,2,\cdots,n\}$. Rich scaling properties of the fluctuations emerge from different view angles \cite{Bak-1996,Sornette-1998-PR,Stanley-1999-RMP,Sornette-2004}. In his seminal work, Taylor documented a power-law relationship between the variance and the mean of populations, which is known as Taylor's law of spatial fluctuation scaling \cite{Taylor-1961-Nature}. Taylor's law has been widely observed in different systems \cite{Eisler-Bartos-Kertesz-2008-AP}, including heavy-ion collisions \cite{Botet-Ploszajczak-Chbihi-Borderie-Durand-Frankland-2001-PRL}, cosmic rays \cite{Uttley-McHardy-2001-MNRAS}, stock trading activities \cite{Eisler-Bartos-Kertesz-2008-AP}, species abundance in population dynamics \cite{Taylor-1961-Nature,Anderson-Gordon-Crawley-Hassell-1982-Nature,Lagrue-Poulin-Cohen-2015-PNAS}, cell numbers \cite{Azevedo-Leroi-2001-PNAS}, hematogenous organ metastases \cite{Kendal-2002-JTB}, human single nucleotide polymorphisms (SNPs) \cite{Kendal-2003-MBE}, individual health \cite{Mitnitski-Rockwood-2006-MAD}, bacterial populations in the human microbiome \cite{Ma-2015-ME}, tornado outbreak \cite{Tippett-Cohen-2016-NC}, and so on. Taylor's law can further be extended to higher orders \cite{Giometto-Formentin-Rinaldo-Cohen-Maritan-2015-PNAS}.

Another form of Taylor's law concerns the temporal fluctuation scaling between the variance and the mean of fluctuation signals. For an observation period $t\in(0,T]$ and an observation resolution $\Delta{t}$, we denote $f_i(\Delta{t},t)$ the fluctuation of constituent $i$ over the time interval $(t-\Delta{t},t]$, whose mean is
\begin{equation}
 m_{\Delta{t}} = \langle{f_i(\Delta{t},t)}\rangle = \frac{1}{J}\sum_{j=1}^{J} f_i(\Delta{t},t),
 \label{Eq:ILL:TFS:mean}
\end{equation}
where $t=j\Delta{t}$ and $J=[T/\Delta{t}]$, and the variance is
\begin{equation}
 V_{\Delta{t}} = \sigma_i^2(\Delta{t}) = \langle{\left[f_i(\Delta{t},t)\right]^2}\rangle-\langle{f_i(\Delta{t},t)}\rangle^2.
 \label{Eq:ILL:TFS:Var}
\end{equation}
Taylor's law of temporal fluctuation scaling reads \cite{Eisler-Bartos-Kertesz-2008-AP}:
\begin{equation}
 V_{\Delta{t}} = a\times(m_{\Delta{t}})^b,
 \label{Eq:ILL:TFS:V:m}
\end{equation}
where $a$ is a positive constant and $b$ is the scaling exponent. To estimate the parameters $a$ and $b$, one can simply rewrite Eq.~(\ref{Eq:ILL:TFS:V:m}) as follows
\begin{equation}
 \ln{V_{\Delta{t}}} = \log{a}+b\log{m_{\Delta{t}}},
 \label{Eq:ILL:TFS:logV:logm}
\end{equation}
where $\log$ is in decimal base, and then performs the ordinary least-squares linear regression.

Taylor's law of temporal fluctuation scaling has also been observed in diverse disciplines \cite{Eisler-Bartos-Kertesz-2008-AP}, such as traffic fluxes recorded at individual nodes in transportation networks (the number of bytes on Internet, the stream flow in river networks, the number of cars on highways) \cite{deMenezes-Barabasi-2004a-PRL,deMenezes-Barabasi-2004b-PRL,Barabasi-deMenezes-Balensiefer-Brockman-2004-EPJB,Duch-Arenas-2006-PRL}, gene expression time series from yeast and human organisms \cite{Nacher-Ochiai-Akutsu-2005-MPLB}, gene network of yeast \cite{Zivkovic-Tadic-Wick-Thurner-2006-EPJB}, trading activities in stock markets \cite{Eisler-Kertesz-Yook-Barabasi-2005-EPL,Kertesz-Eisler-2006,Kertesz-Eisler-2005b-XXX,Eisler-Kertesz-2006-PRE,Eisler-Kertesz-2006-EPJB, Eisler-Kertesz-2006,Eisler-Kertesz-2007-EPL,Eisler-Kertesz-2007-PA, Jiang-Guo-Zhou-2007-EPJB}, application installations \cite{Onnela-ReedTsochas-2010-PNAS}, quotation activities and transaction activities in the foreign exchange market \cite{Sato-Hayashi-Holyst-2012-JEIC}, species abundance \cite{Hekstra-Leibler-2012-Cell,Ma-2015-ME}


We notice that the quantities investigated in the literature of Taylor's law are usually additive, that is,
\begin{equation}
 f_i(\Delta{t},t) = \sum_{t'=t-\Delta{t}+1}^t f_i(1,t').
 \label{Eq:ILL:TFS:additive}
\end{equation}
It is unclear if Taylor's law holds for non-additive quantities. In this paper, we perform a mean-variance analysis on the high-frequency illiquidity time series of Chinese stocks listed on the Shenzhen Stock Exchange (SZSE) and the Shanghai Stock Exchange (SHSE). Illiquidity is the indirect trading cost, reflecting the influence of trading volume on price \cite{Amihud-2002-JFinM}. We confirm the emergence of Taylor's Law of temporal fluctuation scaling in stock illiquidity.

\section{Data description}
\label{S1:Data}

Our data sets contain 2197 A-share and B-Share stocks traded on the Shanghai and Shenzhen Stock Exchanges. The time series have different starting dates since the stocks are gradually listed on the markets and the earliest is in January 1999. The ending dates for most stocks are 30 December 2011 and some other stocks were delisted from the markets.

For each stock $i$, the $\Delta{t}$-min logarithmic returns are calculated from the stock prices $P_i(t)$ as follows,
\begin{equation}
   r_i(\Delta{t},t)=\ln P_i(t)-\ln P_i(t-\Delta{t}).
   \label{Eq:ret}
\end{equation}
We also compute the $\Delta{t}$-min dollar trading volumes $v(t)$ over time intervals $(t-\Delta{t},t]$, which are the sum over all transactions in individual intervals.
The $\Delta{t}$-min illiquidity is defined as the ratio of the absolute $\Delta{t}$-min return to the $\Delta{t}$-min trading volume \cite{Amihud-2002-JFinM}:
\begin{equation}
  f(\Delta{t},t) = \frac{|r(\Delta{t},t)|}{v(\Delta{t},t)}.
  \label{Eq:illiquidity}
\end{equation}
It is clear that that both the returns and the trading volumes are additive, while the illiquidities are non-additive.

The stocks investigate in this work belong to six different markets, that is, the SZSE Main Board (SZMB, with digital codes ``000xxx'' and ``001xxx'') market, the SZSE Small \& Mediate Enterprise Board (SZSMEB, with digital codes ``300xxx'') market, the SZSE Second Board (SZSB, with digital codes ``002xxx'') market, the SHSE Main Board (SHA, with digital codes ``600xxx'') market, the SZSE B-share (SZB, with digital codes ``200xxx'') market, and the SHSE B-share (SHB, with digital codes ``9009xx'') market.

\begin{table}[htb]
  \tbl{Summary statistics of stock illiquidity in different markets. \label{TB:ILL:TFS:SummaryStat}}
{\begin{tabular}{ccccccccccccccccccc}
  \toprule
   Market       &max & mean &median & s.t.d.  &skewness& kurtosis\\
  \colrule
    SZSMEB &  $4.56\times10^{-3}$ & $8.81\times10^{-8}$ & $4.46\times10^{-9}$ & $3.81\times10^{-6}$ & $3.36\times10^2$ & $1.80\times10^5$  \\
      SZSB &  $1.72\times10^{-3}$ & $8.08\times10^{-8}$ & $5.07\times10^{-9}$ & $2.65\times10^{-6}$ & $2.02\times10^2$ & $6.84\times10^4$  \\
      SZMB &  $7.74\times10^{-2}$ & $1.89\times10^{-7}$ & $2.76\times10^{-9}$ & $1.52\times10^{-5}$ & $1.81\times10^3$ & $6.90\times10^6$ \\
       SHA &  $9.21\times10^{-2}$ & $1.28\times10^{-7}$ & $2.14\times10^{-9}$ & $1.35\times10^{-5}$ & $2.85\times10^3$ & $1.64\times10^7$  \\
       SZB &  $1.05$              & $2.20\times10^{-6}$ &                   0 & $3.40\times10^{-4}$ & $2.65\times10^3$ & $8.11\times10^6$  \\
       SHB &  $1.29\times10^{-1}$ & $5.00\times10^{-6}$ & $3.92\times10^{-8}$ & $1.48\times10^{-4}$ & $2.73\times10^2$ & $1.45\times10^5$  \\
  \botrule
\end{tabular}}
  \begin{tabnote}
  The minimum illiquidity for each market is 0.
  \end{tabnote}
\end{table}

Table \ref{TB:ILL:TFS:SummaryStat} presents the summary statistics of 1-min stock illiquidity in the six markets. The B-share stocks (SZB and SHB) have the largest maximum, mean and standard deviation, while the main board A-share stocks (SZMB and SHA) have the second largest maximum, mean and standard deviation. More importantly, the main board stocks (SZMB and SHA) have the smallest median illiquidity, which suggests that on average the main board stocks are more liquid as expected. In contrast, the B-share stocks are the least liquid, which is due to the fact that the two B-share markets are small and the least active. It is also observed that the illiquidities are significantly right-skewed and have fat tails.

\begin{figure}[htb]
  \centerline{\psfig{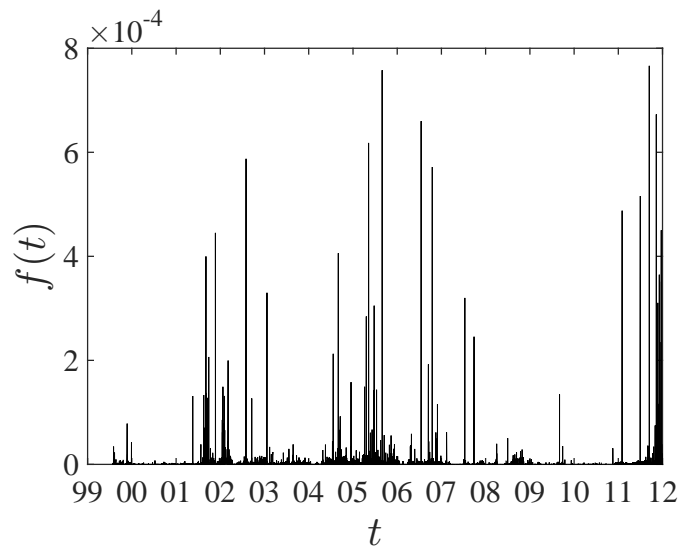}}
  \vspace*{8pt}
  \caption{\label{Fig:ILL:TFS:TS} Evolution of the 1-min illiquidity time series of stock 000016 from 26 July 1999 to 30 December 2011.}
\end{figure}

Figure \ref{Fig:ILL:TFS:TS} illustrates the evolution of the 1-min illiquidity time series of stock 000016 from 26 July 1999 to 30 December 2011. The illiquidity fluctuates remarkably along time and exhibits irregular patterns.

\section{Empirical results}

In order to investigate the relationship between the variance and the mean of stock illiquidity, we compute the illiquidity time series as defined in Eq.~(\ref{Eq:illiquidity}) at different resolutions $\Delta{t}$ for each stock. We then calculate the mean and the variance for each time series. These pairs of mean and variance are used for empirical analyses.

\subsection{Whole sample}
\label{S2:AllStocks}

\begin{figure}[htb]
  \centerline{\psfig{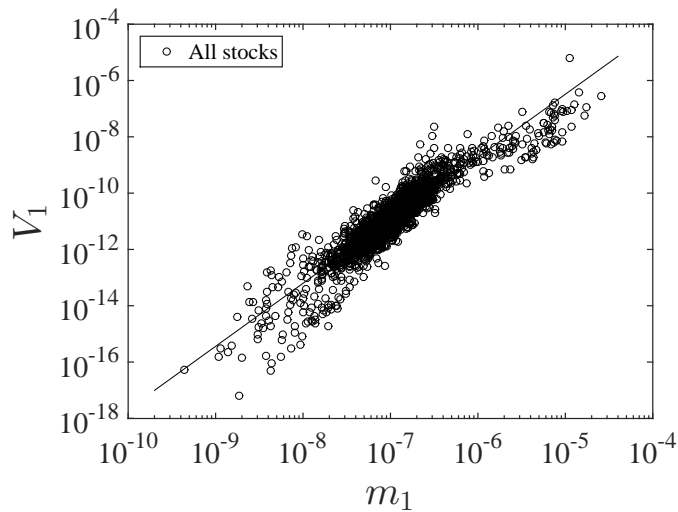}}
  \vspace*{8pt}
  \caption{\label{Fig:ILL:TFS:All} Power-law dependence of the variance $V_1$ on the mean $m_1$ of all the 2197 Chinese stocks.}
\end{figure}

In Fig.~\ref{Fig:ILL:TFS:All}, we show on log-log scales the dependence of the variance $V_1$ on the mean $m_1$ of all the 2197 Chinese stocks, in which the time resolution $\Delta{t}$ is 1 minute. We observe a nice power-law relationship with the scaling range spanning about 5 orders of magnitude. A linear regression of $\log{V_1}$ with respect to $\log{m_1}$ results in $\log{a}=4.70\pm0.14$ and $b=2.24\pm0.02$, which are significantly different from 0 since their $p$-values are almost 0. The adjusted $R^2$ of the fit is 0.86. It is evident that Taylor's law of temporal fluctuation scaling holds for stock illiquidity that is non-additive.

Most of the data points fall within the scope of $[10^{-8},10^{-6}]$ for $m_1$. However, the points corresponding to large illiquidities on the right exhibit remarkable deviations from the power law. According to Table \ref{TB:ILL:TFS:SummaryStat}, B-share stocks have larger illiquidities. A natural conjecture is that these points might belong to B-share markets. Hence, further analyses on Taylor's law of temporal fluctuation scaling should be performed for different ``groups'' of stocks.

\subsection{Different markets}
\label{S2:Markets}

\begin{figure}[htb]
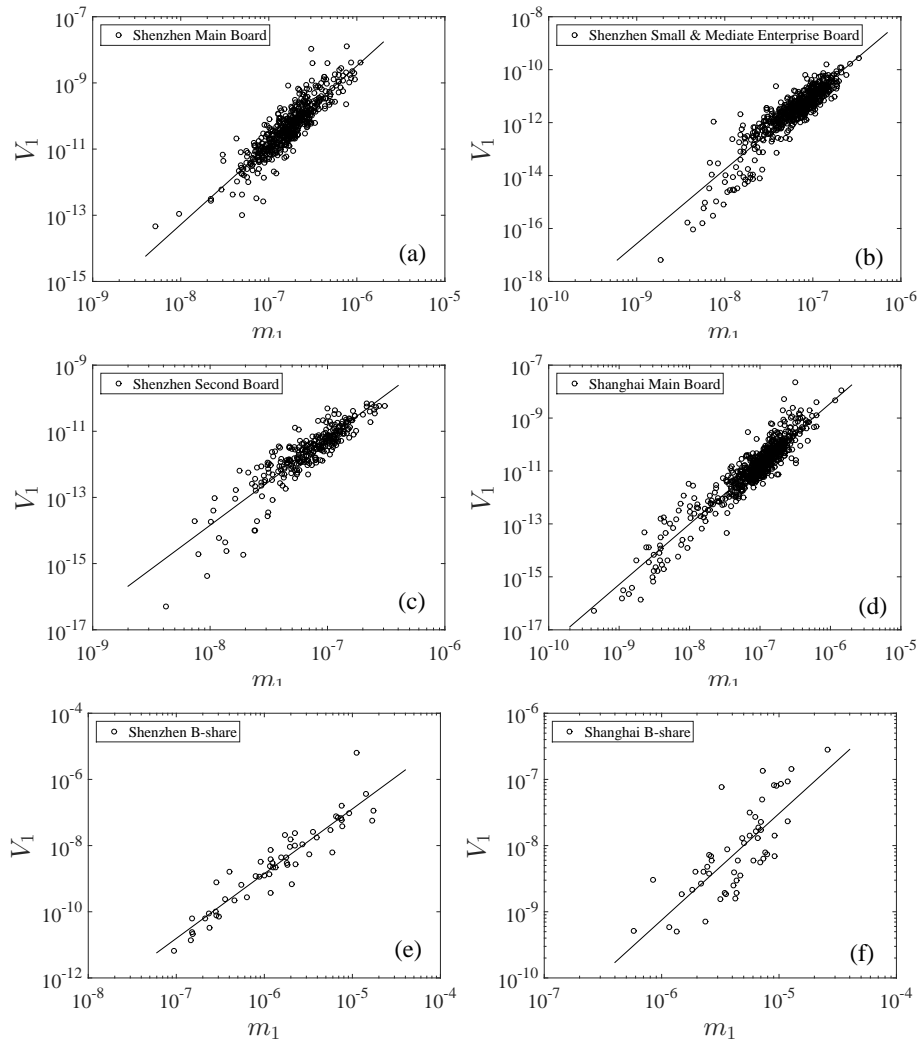

  \centerline{
  \psfig{file=Fig_ILL_TFS_SZMB.eps,width=0.47\linewidth}
  \psfig{file=Fig_ILL_TFS_SZSMEB.eps,width=0.47\linewidth}}
  \centerline{
  \psfig{file=Fig_ILL_TFS_SZSB.eps,width=0.47\linewidth}
  \psfig{file=Fig_ILL_TFS_SHA.eps,width=0.47\linewidth}}
  \centerline{
  \psfig{file=Fig_ILL_TFS_SZB.eps,width=0.47\linewidth}
  \psfig{file=Fig_ILL_TFS_SHB.eps,width=0.47\linewidth}
  }
  \vspace*{8pt}
  \caption{\label{Fig:ILL:TFS:Markets} Power-law dependence of the variance $V_1$ on the mean $m_1$ of the stocks traded in six different Chinese markets.}
\end{figure}

We now perform the mean-variance analysis on the stocks in the six different markets (the SZSE Main Board, the SZSE Small \& Mediate Enterprise Board, the SZSE Second Board, the SHSE Main Board, the SZSE B-share, and the SHSE B-share). The scatter plots between $V_1$ and $m_1$ are illustrated in Fig.~\ref{Fig:ILL:TFS:Markets}. There are evident power-law relationships between the mean and the variance for the whole sample scopes, indicating that Taylor's law holds for different markets. It confirms that the outliers with large $m_1$ in Fig.~\ref{Fig:ILL:TFS:All} come from the data points of the B-share markets as shown in Fig.~\ref{Fig:ILL:TFS:Markets}(e) and (f).

The characteristic parameters of Taylor's law for all the stocks and for the six different markets are obtained by the ordinary least-square linear regression of $\log{V_1}$ against $\log{m_1}$, which are presented in Table \ref{TB:ILL:TFS:Markets}. All the parameters are significantly different from 0. We find that $2<b<3$ for the four A-share markets and $b<2$ for the two B-share markets. This discrepancy might reflect different microstructures of A-share and B-share markets and different behaviors of traders in these two types of markets. A shares are common stocks issued by mainland Chinese companies, subscribed and traded in Chinese Renminbi, listed in mainland Chinese stock exchanges, bought and sold by Chinese nationals. B shares are issued by mainland Chinese companies, traded in foreign currencies and listed in mainland Chinese stock exchanges. B shares carry a face value denominated in Renminbi. The B Share Market was restricted to foreign investors before 19 February 2001 and has been opened to Chinese investors afterwards. In the A-share markets, retail traders dominate and account for about 90\% of the transactions, while in the B-share markets, institutional traders account for about 50\% of the transactions. Possible linkages between traders' behavior and Taylor's law can be studied if trade-level data are available.

\begin{table}[htb]
  \tbl{\label{TB:ILL:TFS:Markets} Summary statistics of illiquidity of 10 stocks. $n$ is the number of stocks in each market. }
{\begin{tabular}{ccccccccccccccccccc}
  \toprule
   Market   & $n$      & $b$  & $p_b$ & $\ln{a}$  & $p_a$  & Adj-$R^2$\\
  \colrule
       All stocks & 2197 & $2.24\pm0.02$ & 0.000 & $4.70\pm0.14$ & 0.000 & 0.86  \\
     SZMB & \hphantom{0}505 & $2.40\pm0.06$ & 0.000 & $5.89\pm0.38$ & 0.000 & 0.79  \\
    SZSMEB & \hphantom{0}646 & $2.80\pm0.05$ & 0.000 & $8.63\pm0.38$ & 0.000 & 0.82  \\
      SZSB & \hphantom{0}281 & $2.64\pm0.08$ & 0.000 & $7.26\pm0.57$ & 0.000 & 0.80  \\
       SHA & \hphantom{0}655 & $2.28\pm0.03$ & 0.000 & $5.27\pm0.24$ & 0.000 & 0.87  \\
       SZB & \hphantom{00}57 & $1.96\pm0.10$ & 0.000 & $2.89\pm0.57$ & 0.000 & 0.88  \\
       SHB & \hphantom{00}53 & $1.61\pm0.18$ & 0.000 & $0.53\pm0.98$ & 0.590 & 0.60  \\
  \botrule
\end{tabular}}
  \begin{tabnote}
  The variable $n$ is the number of stocks in each case and $b$ is the scaling exponent of Taylor's law.
  \end{tabnote}
\end{table}

\subsection{Different industry categories}
\label{S2:Category}

According to the ``{\textit{Guidelines for the Industry Classification of Listed Companies}} (2012 Revision)'' of the China Securities Regulation Commission, Chinese companies are classified into 19 industry categories containing 90 industries. Here we investigate Taylor's law for the stocks in different industry categories. The category codes are alphabets: A for agriculture, forestry, animal husbandry and fishery, B for mining industry, C for manufacturing industry, D for industry of electric power, heat, gas and water production and supply, E for construction industry, F for wholesale and retail industry, G for transport, storage and postal service industry, H for accommodation and catering industry, I for industry of information transmission, software and information technology services, J for financial industry, K for real estate industry, L for leasing and commercial service industry, M for scientific research and technical service industry, N for water conservancy, environment and public facility management industry, O for industry of resident service, repair and other services, P for education, Q for health and social work, R for industry of culture, sports and entertainment, and S for diversified industries. For the 2197 stocks in our sample, categories O and P are not present. It is not surprising that Taylor's law holds for different industry categories. The $b$ values are listed in the first column of Table \ref{TB:ILL:TFS:Category}. It shows that $b\in(2,3)$ in all cases.

\setlength\tabcolsep{1.5pt}
\begin{table}[htb]
  \tbl{\label{TB:ILL:TFS:Category} Regression results of Taylor's temporal fluctuation scaling law in the 1-min illiquidity of all A-share stocks in different categories.}
  {\begin{tabular}{ccccccccccccccccccc}
  \toprule
   && \multicolumn{2}{c}{All stocks} && \multicolumn{2}{c}{SZ MB} && \multicolumn{2}{c}{SZ SMEB} && \multicolumn{2}{c}{SZ SB} && \multicolumn{2}{c}{SHA}\\
  \cline{3-4}\cline{6-7}\cline{9-10}\cline{12-13}\cline{15-16}
  Category && $n$ & $b$ && $n$ & $b$ && $n$ & $b$ && $n$ & $b$ && $n$ & $b$\\
  \colrule
     A &&  34 & $2.75\pm0.10$ &&   5 & $2.80\pm0.35$ &&  11 & $2.42\pm0.36$ &&   5 & $3.54\pm1.12$ &&  13 & $2.80\pm0.15$   \\
     B &&  63 & $2.26\pm0.08$ &&  19 & $2.30\pm0.13$ &&   5 & $2.51\pm1.11$ &&   4 & $2.61\pm0.79$ &&  35 & $2.36\pm0.11$   \\
     C && 1292 & $2.73\pm0.04$ && 247 & $2.31\pm0.08$ && 515 & $2.85\pm0.06$ && 201 & $2.78\pm0.10$ && 329 & $2.27\pm0.07$   \\
     D &&  67 & $2.26\pm0.12$ &&  30 & $2.78\pm0.23$ &&   3 & $2.79\pm0.80$ &&   0 & / &&  34 & $1.88\pm0.17$   \\
     E &&  55 & $2.24\pm0.18$ &&   7 & $2.45\pm0.76$ &&  22 & $2.57\pm0.28$ &&   4 & $2.70\pm1.29$ &&  22 & $2.13\pm0.29$   \\
     F &&  96 & $2.56\pm0.13$ &&  37 & $3.15\pm0.36$ &&  17 & $2.17\pm0.33$ &&   4 & $1.86\pm0.09$ &&  38 & $2.30\pm0.17$   \\
     G &&  63 & $2.09\pm0.11$ &&  13 & $1.96\pm0.19$ &&   6 & $2.15\pm0.47$ &&   2 & / &&  42 & $2.17\pm0.13$   \\
     H &&  10 & $2.58\pm0.18$ &&   6 & $2.09\pm0.29$ &&   2 & / &&   0 & / &&   2 & /   \\
     I &&  94 & $2.11\pm0.15$ &&   8 & $1.84\pm0.40$ &&  34 & $2.79\pm0.24$ &&  41 & $2.08\pm0.19$ &&  11 & $1.22\pm0.34$   \\
     J &&  38 & $2.35\pm0.12$ &&   9 & $2.13\pm0.29$ &&   2 & / &&   0 & / &&  27 & $2.52\pm0.16$   \\
     K &&  99 & $2.37\pm0.10$ &&  53 & $2.52\pm0.15$ &&   9 & $1.81\pm0.54$ &&   0 & / &&  37 & $2.43\pm0.17$   \\
     L &&  20 & $2.32\pm0.30$ &&   6 & $3.33\pm1.30$ &&   6 & $2.56\pm0.56$ &&   4 & $3.12\pm0.72$ &&   4 & $2.17\pm0.10$   \\
     M &&   8 & $2.87\pm0.29$ &&   0 & / &&   4 & $3.15\pm0.24$ &&   4 & $2.29\pm0.62$ &&   0 & /   \\
     N &&  20 & $2.56\pm0.23$ &&   8 & $2.49\pm0.58$ &&   6 & $2.42\pm0.55$ &&   4 & $1.54\pm0.25$ &&   2 & /   \\
     Q &&   3 & $2.42\pm0.16$ &&   0 & / &&   1 & / &&   2 & / &&   0 & /   \\
     R &&  26 & $2.54\pm0.16$ &&   9 & $1.80\pm0.31$ &&   3 & $4.01\pm0.81$ &&   5 & $2.54\pm0.46$ &&   9 & $2.18\pm0.21$   \\
     S &&  15 & $2.28\pm0.38$ &&   7 & $2.09\pm0.54$ &&   0 & / &&   0 & / &&   8 & $2.35\pm0.58$   \\
  \botrule
  \end{tabular}}
  \begin{tabnote}
  The variable $n$ is the number of stocks in each case and $b$ is the scaling exponent of Taylor's law.
  \end{tabnote}
\end{table}

We further investigate the impact of markets. Because there are only 57 SZB stocks and 53 SHB stocks (see Table \ref{TB:ILL:TFS:Markets}), these two markets are excluded from our analysis. The estimated scaling exponents $b$ are listed in Table \ref{TB:ILL:TFS:Category}. Although most of the scaling exponents fall within $(2,3)$, some are less than 2 and others are greater than 3. With less stocks in individual markets (SZMB, SZSMEB, SZSB and SHA) than the case of all stocks, the scaling exponents exhibit larger fluctuations. When the number of stocks is too small, it is meaningless to perform the mean-variance analysis.

\subsection{Different sectors}
\label{S2:Sector}

In the Chinese stock markets, there are different classification standards for stock industries. Although the majority of them are similar, there are also differences in the naming and classification of stocks. According to the Shenwan Industry Classification Standard, a widely used standard, the stocks under investigation belong to 28 sectors: Agriculture, Architectural ornament, Automobile, Bank, Building material, Catering \& tourism, Chemical, Commerce \& trade, Computer, Electrical equipment, Electrical household appliances, Electronic component, Ferrous metal, Financial, Food \& beverage, Light industry manufacturing, Mechanical equipment, Medias, Medical biology, Mining, Miscellaneous, Nonferrous metal, Property, Public utility, Telecommunication, Textile \& clothing, Transportation \& infrastructure, and War industry.

We perform mean-variance analyses on stocks belonging to different sectors. Table \ref{TB:ILL:TFS:Sectors} shows the scaling exponents of stocks in different sectors. Again, most scaling exponents $b\in(2,3)$. There are also cases with the scaling exponents $b<2$ or $b>3$.

\setlength\tabcolsep{1.5pt}
\begin{table}[htb]
  \tbl{\label{TB:ILL:TFS:Sectors}Regression results of Taylor's temporal fluctuation scaling law in the 1-min illiquidity of all A-share stocks in different sectors.}
  {\begin{tabular}{ccccccccccccccccccc}
  \toprule
   && \multicolumn{2}{c}{All stocks} && \multicolumn{2}{c}{SZ MB} && \multicolumn{2}{c}{SZ SMEB} && \multicolumn{2}{c}{SZ SB} && \multicolumn{2}{c}{SHA}\\
  \cline{3-4}\cline{6-7}\cline{9-10}\cline{12-13}\cline{15-16}
  Sector && $n$ & $b$ && $n$ & $b$ && $n$ & $b$ && $n$ & $b$ && $n$ & $b$\\
  \colrule
   Transportation \& Infra. &&  69 & $2.11\pm0.11$ &&  13 & $1.96\pm0.19$ &&   8 & $2.12\pm0.39$ &&   2 & / &&  46 & $2.18\pm0.12$   \\
   Catering \& Tourism &&  25 & $2.39\pm0.16$ &&  11 & $2.61\pm0.38$ &&   6 & $2.32\pm0.50$ &&   2 & / &&   6 & $1.76\pm0.41$   \\
       Medias &&  48 & $2.51\pm0.14$ &&  11 & $2.27\pm0.24$ &&  15 & $3.03\pm0.44$ &&  13 & $2.20\pm0.22$ &&   9 & $2.37\pm0.31$   \\
   Public utility &&  85 & $2.45\pm0.12$ &&  32 & $2.78\pm0.23$ &&   5 & $2.64\pm0.50$ &&  13 & $1.64\pm0.19$ &&  35 & $1.94\pm0.18$   \\
   Agriculture &&  69 & $2.74\pm0.09$ &&  15 & $2.81\pm0.21$ &&  21 & $2.45\pm0.28$ &&  11 & $1.97\pm0.44$ &&  22 & $2.79\pm0.12$   \\
     Chemical && 206 & $2.88\pm0.09$ &&  46 & $2.64\pm0.39$ &&  82 & $3.08\pm0.12$ &&  30 & $2.49\pm0.21$ &&  48 & $2.00\pm0.25$   \\
   Medical Biology && 147 & $2.65\pm0.10$ &&  33 & $2.48\pm0.31$ &&  41 & $2.26\pm0.19$ &&  32 & $2.46\pm0.30$ &&  41 & $2.53\pm0.16$   \\
   Commerce \& Trade &&  53 & $2.54\pm0.17$ &&  21 & $3.59\pm0.44$ &&  12 & $1.74\pm0.30$ &&   1 & / &&  19 & $2.42\pm0.38$   \\
   War Industry &&  27 & $1.97\pm0.18$ &&   2 & / &&   6 & $2.09\pm0.51$ &&   3 & $-1.16\pm0.15$ &&  16 & $1.92\pm0.19$   \\
   Electr. Household Appl. &&  38 & $3.19\pm0.25$ &&  10 & $1.52\pm0.62$ &&  20 & $3.36\pm0.34$ &&   5 & $5.53\pm0.34$ &&   3 & $2.19\pm0.58$   \\
   Building Material &&  50 & $2.91\pm0.15$ &&  12 & $2.35\pm0.29$ &&  23 & $3.21\pm0.30$ &&   4 & $2.31\pm1.86$ &&  11 & $2.37\pm0.27$   \\
   Architectural Ornament &&  52 & $2.28\pm0.19$ &&   5 & $2.39\pm0.87$ &&  25 & $2.56\pm0.26$ &&   1 & / &&  21 & $2.23\pm0.30$   \\
     Property && 105 & $2.33\pm0.10$ &&  59 & $2.47\pm0.15$ &&   8 & $2.01\pm0.60$ &&   0 & / &&  38 & $2.36\pm0.18$   \\
   Nonferrous Metal &&  84 & $2.55\pm0.08$ &&  25 & $2.41\pm0.24$ &&  22 & $2.84\pm0.20$ &&   2 & / &&  35 & $2.47\pm0.12$   \\
   Mechanical Equipment && 171 & $2.75\pm0.11$ &&  26 & $2.44\pm0.27$ &&  66 & $3.06\pm0.20$ &&  44 & $3.14\pm0.18$ &&  35 & $2.11\pm0.17$   \\
   Automobile &&  77 & $2.63\pm0.11$ &&  27 & $2.21\pm0.13$ &&  27 & $3.00\pm0.27$ &&   4 & $2.81\pm1.56$ &&  19 & $2.51\pm0.24$   \\
   Electronic Component && 112 & $3.12\pm0.14$ &&  16 & $2.31\pm0.33$ &&  53 & $3.23\pm0.20$ &&  29 & $3.12\pm0.27$ &&  14 & $3.03\pm0.82$   \\
   Electrical Equipment && 113 & $2.47\pm0.13$ &&   6 & $1.51\pm0.37$ &&  53 & $2.43\pm0.22$ &&  23 & $2.19\pm0.29$ &&  31 & $2.42\pm0.19$   \\
   Textile \& Clothing &&  60 & $2.59\pm0.17$ &&   7 & $2.12\pm0.54$ &&  34 & $2.31\pm0.25$ &&   1 & / &&  18 & $2.68\pm0.24$   \\
   Miscellaneous &&  33 & $1.96\pm0.17$ &&  12 & $1.68\pm0.25$ &&   1 & / &&   1 & / &&  19 & $2.17\pm0.25$   \\
     Computer &&  93 & $2.35\pm0.14$ &&   8 & $2.34\pm0.43$ &&  37 & $2.51\pm0.30$ &&  39 & $2.12\pm0.20$ &&   9 & $2.38\pm0.15$   \\
   Light Ind Manufact &&  67 & $2.84\pm0.18$ &&  10 & $2.60\pm0.42$ &&  37 & $2.80\pm0.24$ &&   3 & $2.86\pm0.31$ &&  17 & $2.11\pm0.41$   \\
   Telecommunication &&  51 & $1.91\pm0.24$ &&   8 & $1.49\pm0.62$ &&  18 & $3.10\pm0.31$ &&  13 & $1.71\pm0.46$ &&  12 & $1.25\pm0.37$   \\
       Mining &&  49 & $2.19\pm0.10$ &&  13 & $2.29\pm0.19$ &&   4 & $3.48\pm1.09$ &&   3 & $2.55\pm1.12$ &&  29 & $2.22\pm0.12$   \\
   Ferrous Metal &&  30 & $2.09\pm0.29$ &&  10 & $1.60\pm0.40$ &&   5 & $1.27\pm1.16$ &&   0 & / &&  15 & $2.31\pm0.43$   \\
         Bank &&  16 & $2.62\pm0.50$ &&   1 & / &&   1 & / &&   0 & / &&  14 & $2.62\pm0.55$   \\
    Financial &&  23 & $2.30\pm0.14$ &&   9 & $2.39\pm0.73$ &&   1 & / &&   0 & / &&  13 & $2.48\pm0.18$   \\
   Food \& Beverage &&  50 & $2.92\pm0.16$ &&  16 & $2.64\pm0.24$ &&  15 & $2.89\pm0.31$ &&   1 & / &&  18 & $2.33\pm0.37$   \\
  \botrule
  \end{tabular}}
  \begin{tabnote}
  The variable $n$ is the number of stocks in each case and $b$ is the scaling exponent of Taylor's law.
  \end{tabnote}
\end{table}

\subsection{Different provinces}
\label{S2:Prov}

We also investigate stocks in 31 different provinces and direct-controlled municipalities (Beijing, Shanghai, Tianjin and Chongqing). The results are shown in Table \ref{TB:ILL:TFS:Prov}. We find that most of the scaling exponents are greater than 2 and less than 3 with a few exceptions. There is a negative scaling exponent for three Chongqing companies which are listed on the Small \& Mediate Enterprise Board of the Shenzhen Stock Exchange.

\setlength\tabcolsep{1.5pt}

\begin{table}[htb]
  \tbl{\label{TB:ILL:TFS:Prov} Regression results of Taylor's temporal fluctuation scaling law in the 1-min illiquidity of all A-share stocks in different provinces.}
  {\begin{tabular}{ccccccccccccccccccc}
  \toprule
  Province/ && \multicolumn{2}{c}{All stocks} && \multicolumn{2}{c}{SZ MB} && \multicolumn{2}{c}{SZ SMEB} && \multicolumn{2}{c}{SZ SB} && \multicolumn{2}{c}{SHA}\\
  \cline{3-4}\cline{6-7}\cline{9-10}\cline{12-13}\cline{15-16}
  Municipality && $n$ & $b$ && $n$ & $b$ && $n$ & $b$ && $n$ & $b$ && $n$ & $b$\\
  \colrule
     Shanghai &&  96 & $2.28\pm0.12$ &&   2 & / &&  26 & $2.79\pm0.27$ &&  23 & $2.58\pm0.30$ &&  45 & $2.30\pm0.14$   \\
       Yunnan &&  25 & $2.51\pm0.16$ &&   8 & $2.43\pm0.49$ &&   7 & $2.50\pm0.31$ &&   1 & / &&   9 & $2.54\pm0.25$   \\
   Inner Mongolia &&  21 & $1.70\pm0.32$ &&   5 & $0.85\pm1.56$ &&   1 & / &&   2 & / &&  13 & $1.68\pm0.47$   \\
      Beijing && 177 & $2.14\pm0.08$ &&  28 & $2.09\pm0.18$ &&  32 & $2.23\pm0.26$ &&  39 & $2.47\pm0.26$ &&  78 & $2.23\pm0.09$   \\
        Jilin &&  33 & $2.85\pm0.16$ &&  14 & $2.56\pm0.19$ &&   6 & $3.50\pm0.82$ &&   1 & / &&  12 & $2.48\pm0.26$   \\
      Sichuan &&  72 & $2.78\pm0.11$ &&  23 & $2.26\pm0.16$ &&  21 & $3.10\pm0.28$ &&   7 & $2.65\pm0.64$ &&  21 & $2.61\pm0.27$   \\
      Tianjin &&  30 & $2.05\pm0.18$ &&   7 & $2.96\pm1.18$ &&   6 & $2.41\pm0.35$ &&   4 & $2.21\pm1.43$ &&  13 & $1.93\pm0.25$   \\
      Ningxia &&  11 & $2.75\pm0.27$ &&   7 & $2.77\pm0.65$ &&   1 & / &&   0 & / &&   3 & $1.58\pm0.27$   \\
        Anhui &&  74 & $2.88\pm0.14$ &&  16 & $2.88\pm0.23$ &&  24 & $3.39\pm0.30$ &&   7 & $3.17\pm0.23$ &&  27 & $2.23\pm0.24$   \\
     Shandong && 125 & $2.65\pm0.10$ &&  28 & $2.09\pm0.14$ &&  54 & $2.74\pm0.19$ &&  14 & $2.91\pm0.33$ &&  29 & $2.14\pm0.13$   \\
       Shanxi &&  31 & $2.45\pm0.19$ &&  11 & $2.40\pm0.24$ &&   3 & $2.47\pm0.15$ &&   2 & / &&  15 & $2.19\pm0.22$   \\
    Guangdong && 319 & $2.50\pm0.06$ &&  93 & $2.34\pm0.11$ && 132 & $2.84\pm0.11$ &&  61 & $2.68\pm0.16$ &&  33 & $2.16\pm0.09$   \\
      Guangxi &&  27 & $1.94\pm0.39$ &&  12 & $1.93\pm0.50$ &&   5 & $1.81\pm0.59$ &&   0 & / &&  10 & $1.94\pm0.88$   \\
      Xijiang &&  34 & $1.73\pm0.22$ &&   5 & $2.94\pm1.15$ &&   8 & $0.84\pm0.28$ &&   2 & / &&  19 & $2.33\pm0.32$   \\
      Jiangsu && 188 & $2.88\pm0.10$ &&  23 & $2.10\pm0.48$ &&  87 & $2.89\pm0.16$ &&  28 & $3.30\pm0.23$ &&  50 & $2.39\pm0.16$   \\
      Jiangxi &&  27 & $2.51\pm0.12$ &&   7 & $2.94\pm0.38$ &&   6 & $2.52\pm0.14$ &&   2 & / &&  12 & $2.33\pm0.17$   \\
        Hebei &&  41 & $2.40\pm0.20$ &&  14 & $2.03\pm0.32$ &&   9 & $2.74\pm0.45$ &&   5 & $2.52\pm0.19$ &&  13 & $2.44\pm0.42$   \\
        Henan &&  58 & $2.85\pm0.13$ &&  10 & $2.50\pm0.51$ &&  22 & $3.17\pm0.26$ &&   8 & $2.15\pm0.16$ &&  18 & $2.52\pm0.19$   \\
     Zhejiang && 203 & $2.98\pm0.09$ &&  13 & $2.42\pm0.50$ && 113 & $3.04\pm0.11$ &&  26 & $2.56\pm0.30$ &&  51 & $2.95\pm0.25$   \\
       Hainan &&  24 & $2.78\pm0.16$ &&  13 & $2.57\pm0.48$ &&   2 & / &&   2 & / &&   7 & $2.87\pm0.11$   \\
        Hubei &&  72 & $2.96\pm0.20$ &&  26 & $2.42\pm0.38$ &&   9 & $2.03\pm0.43$ &&  10 & $3.59\pm0.50$ &&  27 & $2.59\pm0.42$   \\
        Hunan &&  61 & $2.27\pm0.12$ &&  22 & $2.56\pm0.24$ &&  17 & $1.74\pm0.17$ &&   8 & $1.54\pm0.41$ &&  14 & $2.38\pm0.30$   \\
        Gansu &&  20 & $2.66\pm0.26$ &&   7 & $2.22\pm0.68$ &&   4 & $4.70\pm0.80$ &&   2 & / &&   7 & $3.35\pm1.16$   \\
       Fujian &&  68 & $2.49\pm0.11$ &&  14 & $2.45\pm0.31$ &&  27 & $2.79\pm0.28$ &&   9 & $2.39\pm0.36$ &&  18 & $2.41\pm0.18$   \\
        Tibet &&   6 & $2.80\pm0.40$ &&   2 & / &&   1 & / &&   0 & / &&   3 & $2.42\pm0.69$   \\
      Guizhou &&  17 & $2.91\pm0.38$ &&   5 & $3.76\pm1.63$ &&   5 & $2.04\pm0.14$ &&   0 & / &&   7 & $3.00\pm0.39$   \\
     Liaoning &&  51 & $2.37\pm0.16$ &&  20 & $1.77\pm0.54$ &&  10 & $2.31\pm0.30$ &&   6 & $2.21\pm0.07$ &&  15 & $2.22\pm0.27$   \\
    Chongqing &&  31 & $2.91\pm0.21$ &&  11 & $2.95\pm0.29$ &&   3 & $-1.44\pm4.19$ &&   4 & $5.47\pm0.78$ &&  13 & $2.62\pm0.33$   \\
      Shaanxi &&  33 & $2.11\pm0.15$ &&  11 & $2.79\pm0.61$ &&   3 & $1.67\pm0.10$ &&   6 & $2.06\pm0.55$ &&  13 & $2.23\pm0.20$   \\
      Qinghai &&   7 & $3.29\pm0.29$ &&   2 & / &&   1 & / &&   0 & / &&   4 & $3.12\pm0.33$   \\
   Heilongjiang &&  21 & $2.77\pm0.19$ &&   5 & $2.47\pm0.52$ &&   1 & / &&   1 & / &&  14 & $2.97\pm0.24$   \\
  \botrule
  \end{tabular}}
  \begin{tabnote}
  In the first column, Beijing, Shanghai, Tianjin and Chongqing are direct-controlled municipalities, while others are provinces. The variable $n$ is the number of stocks in each case and $b$ is the scaling exponent of Taylor's law.
  \end{tabnote}
\end{table}

\subsection{Dependence of $\Delta{t}$}
\label{S2:Dt}

\begin{figure}[htb]
  \centerline{\psfig{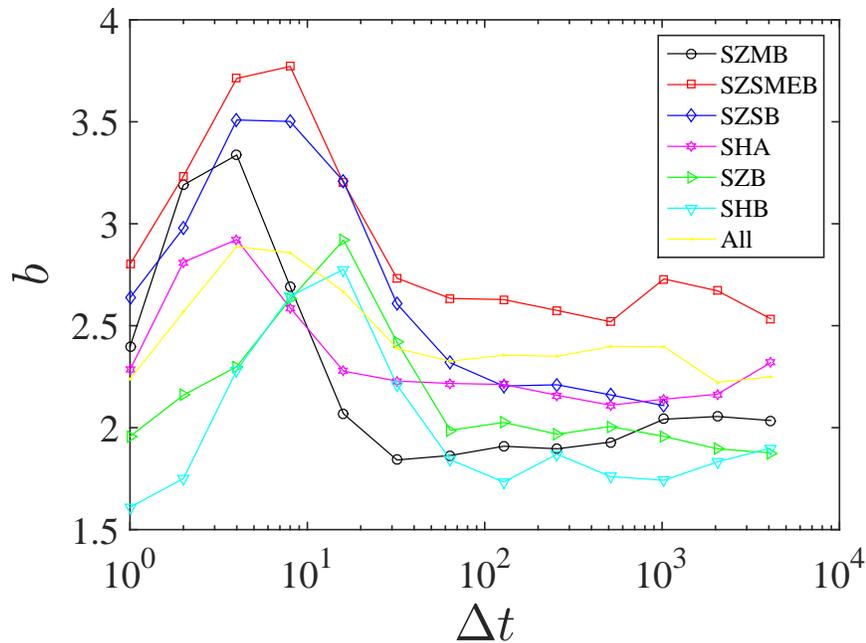}}
  \vspace*{8pt}
  \caption{\label{Fig:ILL:TFS:b:Dt} Dependence of the scaling exponent $b$ on the time interval $\Delta{t}$.}
\end{figure}

In the previous subsections, we focused on the 1-min illiquidity data, that is, $\Delta{t}=1$ min. We find that, for larger intervals, Taylor's law presented in Eq.~(\ref{Eq:ILL:TFS:V:m}) also holds for the whole sample and the six markets. Figure \ref{Fig:ILL:TFS:b:Dt} illustrates the dependence of the scaling exponent $b$ on the time interval $\Delta{t}$ for the seven cases. For small $\Delta{t}$ values, there is a rough logarithmic relationship
\begin{equation}
  b \sim \log\Delta{t}.
  \label{Eq:ILL:TFS:b:Dt}
\end{equation}
This logarithmic relationship was observed over much wider intervals for many additive quantities \cite{Eisler-Kertesz-2006-PRE,Eisler-Kertesz-2007-PA,Jiang-Guo-Zhou-2007-EPJB,Eisler-Bartos-Kertesz-2008-AP}.

With the increase of $\Delta{t}$, the exponent $b$ reaches its maximum soon at $\Delta{t}_{\max}$, which is the smallest for the two main board markets (SZMB and SHA) and the largest for the two B-share markets (SZB and SHB). After the peak for each market, the exponent $b$ decreases rapidly to a stable level. On average, the ``asymptotic'' scaling exponent is larger than 2 for the four A-share markets and less than 2 for the two B-share markets.

\section{Conclusion}
\label{S1:con}

In summary, we have documented significant evidence of Taylor's law of temporal fluctuation scaling in the illiquidity of Chinese stocks. Taylor's law holds for stocks in different markets, in different industry categories, in different sectors, and in different regions. However, there is no evidence showing that this scaling law is universal, since the scaling exponents fluctuate from groups to groups. In particular, the scaling exponent is less than 2 for B-share markets and greater than 2 for A-share markets.

Fr further research, it would be interesting to identify factors that might impact the value of $b$. Possible factors contain the capitalization of stocks, the investor diversity of stocks, or the GDP of the considered Chinese provinces on the aggregate level.

In literature, quantities that comply with Taylor's law are usually additive. Hence, one of the main contributions of this work is that Taylor's law of temporal fluctuation scaling holds for non-additive stock illiquidity investigated in this work. We conjecture that Taylor's law might also hold for other non-additive quantities in other complex systems.

However, due to the non-additivity of illiquidity, we also unveiled intriguing different behaviors. Specifically, the scaling exponent is observed to increase with $\Delta{t}$ in a logarithmical form and reach its maximum soon, after which the exponent decreases rapidly and remains at a stable level. The stable level differs for stocks in different markets.

\section*{Acknowledgements}

The data sets used in this work were kindly provided by a leading Chinese stock market data provider RESSET (http://resset.cn/) for scientific research. This work was partly supported by the National Natural Science Foundation of China (Grants No. 71501072, 71532009 and 71131007) and the Fundamental Research Funds for the Central Universities.


\end{document}